\documentstyle[epsf,bibnorm]{lamuphys}
\makeatletter
\let\chapter\hid@chapter
\makeatother

\newcommand{\bce}{\begin{center}}
\newcommand{\ece}{\end{center}}
\newcommand{\beq}{\begin{equation}}
\newcommand{\eeq}{\end{equation}}
\newcommand{\bea}{\begin{eqnarray}}
\newcommand{\eea}{\end{eqnarray}}

\def\eps{\varepsilon}

\def\om{\omega}

\def\N{{\scriptstyle N}}
\begin{document}
\pagenumbering{arabic}

\title{Nuclear matter, nuclear and subnuclear degrees of freedom}

\author{Wanda\,M.\,Alberico\inst{1,2} 
}

\institute{ Dipartimento di Fisica Teorica, Universit\`a di Torino,
via P. Giuria 1,\, I--10125 Torino, Italy
\and
INFN, Sezione di Torino, Italy
 }

\maketitle

\begin{abstract}
We report here  theoretical investigations on the  complexity 
of nuclear structure, which have been carried out in the framework
of different many--body approaches, typically applied to nuclear
matter and quark matter studies. The variational, functional and 
perturbative scheme are illustrated in their latest developments.
The effect of various nucleon--nucleon interactions are tested, 
particularly in the context of the nuclear response functions, 
against a large body of experimental data. The properties and 
decay widths of hypernuclei are shortly revisited, while the 
equation of state of  isospin asymmetric nuclear matter leads us
toward nuclear systems of astrophysical interest, like the neutron
stars. Finally the transition from hadronic matter to the deconfined
phase of quark--gluon plasma endorses the application of many--body 
and field theoretical techniques to a system with subnucleonic 
degrees of freedom.
\end{abstract}

\section{Introduction}

The subject of this report  covers a wide range of items and problems 
in the theoretical description of nuclear systems, the common link 
among them being the strong interaction intervening between
 the constituents of composite systems and the many--body complexity
of the latter.

Nuclear and nucleonic structure has been experimentally investigated 
using probes of quite different nature, whose interaction with the 
nuclear system  involves forces of different nature, from the 
electroweak one to the most complicated strong interactions between 
relativistic heavy ions. 

We will discuss here at length the information which can be 
obtained through unpolarized and polarized electron--nucleus 
scattering, briefly touching also neutrino scattering. 
Strongly interacting probes, like mesons ($\pi, K$) and nucleons will 
be considered in connection with the specific spin--isospin channels 
they can excite in a nuclear system. 

Finally we will briefly illustrate the problems connected with 
subnucleonic degrees of freedom, in particular the transition from 
nuclear matter to quark matter:  the so--called deconfinement 
phase transition, perhaps the only one occurring in the early stages of 
the Universe which one can hope to reproduce in the laboratory. 

Different theoretical schemes are employed to test the performances 
and limitations of various models for describing nuclear dynamics and 
spectra. Indeed in the nuclear medium many degrees of freedom come 
into play and require different techniques to be enlightened.

For example, sticking to the case of electromagnetic probes, one has
to consider not only the one--body nucleonic current, but also the
two--body ones, which involve pairs of nucleons correlated via 
the  exchange of charged mesons. In the domain of deep inelastic 
scattering the high momentum photons exchanged between the electron
and the nucleus shed light on the quark structure of the nucleon, 
namely on  the complications associated with the 
subnucleonic degrees of freedom and colour confinement. 

Just mentioning all the above items and themes can give the 
impression of many different, disconnected fields of research: on 
the contrary, theoretical and experimental nuclear physics can
be viewed as a ``magic circle'' which involves, with various and 
strong interrelations, all different subfields. One can start, 
e.g., from the single nucleon to proceed toward the structure of
nuclei and hypernuclei, down to the ideal ``nuclear matter'' system, 
then to neutron stars and astrophysical systems, in which realistic 
models envisage the existence of a quark--gluon plasma phase. And 
then back, through the confinement mechanism, to nucleons and 
nuclear systems. 

\section{Weakly interacting probes: electrons and neutrinos}

The main advantage of electron--nucleus scattering stems from the 
fairly good knowledge of the basic electromagnetic interaction vertex;
moreover the virtual exchanged photon can penetrate well inside the 
nuclear system before being absorbed, thus involving into the 
process any nuclear constituent in the whole volume of the nucleus. 
This probe allows to investigate, depending upon the four--momentum 
transfer, both {\it nuclear} and {\it nucleonic} structure. In the 
second case, which concerns deep inelastic scattering experiments, 
the partonic structure and perturbative QCD corrections can be tested,
together with various quark models for the confined hadronic states. 

In the former case, which refers to relatively smaller momentum transfers,
nuclear structure studies can be performed by comparing the theoretical 
evaluation of the inelastic electron scattering cross sections with the 
measured data, in a wide range of energy and momentum transfers as well 
as in different experimental setups, appropriate for inclusive $(e,e')$, 
one--exclusive $(e,e'p)$, etc. measurements.

In all situations we  test the theoretical description of complex nuclei 
in connection with both the many--body scheme employed and the role played 
by the NN interaction in shaping cross sections and  amplitudes 
for the excitation of specific nuclear states. 
Taking advantage from the fact that electrons interact with the whole 
nuclear volume, one expects that surface effects are of minor importance:
thus it is convenient to perform calculations in the framework of nuclear 
matter. 

Both perturbative and variational techniques have been employed to 
test different model descriptions of:
\begin{enumerate}
\item 
  the two (and three)--body nucleon--nucleon interaction
\item 
  the nuclear electromagnetic currents and form factors 
(including the single nucleon current as well as the two--body, 
meson exchange currents)
\item 
  the weak nuclear and nucleonic currents (e.g. to investigate 
the strange form factors of the nucleon)
\item 
  relativistic effects both in the kinematics and in the currents.
\end{enumerate}
All the above items concur in determining the so--called 
{\bf nuclear response functions}, which can be directly confronted with 
the ones extracted from the experimental data.

Different types of response functions can be obtained  from the cross 
sections for inelastic electron scattering; we shall list below a few 
items which have been extensively explored.

\begin{description}
\item[$\diamondsuit$] 
{\sl Inclusive scattering of unpolarized electrons off nuclei}

In Born approximation the differential cross section (with respect to
the energy, $\epsilon'$, and the scattering angles, 
$\Omega\equiv(\theta,\varphi)$, of the final electron) is given by:
\beq
\frac{d^2\sigma}{d\Omega d\epsilon'} = 
\sigma_M \left\{ v_L R_L(q,\omega) 
+ v_T R_T(q,\omega)\right \}
\label{cross}
\eeq
where $\sigma_M$ is the Mott cross section, $q_{\mu}^2=\omega^2-q^2$ 
the squared four--momentum transfer, $v_L=(q_{\mu}^2/q^2)^2$, 
$v_T=\left[\tan^2\left(\theta/2\right) - q_{\mu}^2/2q^2\right]$ and
$R_{L,T}$ the longitudinal and transverse (separated) electromagnetic 
nuclear response functions. Relevant physical issues are connected with 
these response functions, in particular the so--called {\it Coulomb 
sum rule} (from the longitudinal response integrated over the energy)
and the scaling properties, both in non--relativistic and in relativistic
regimes.

\item[$\diamondsuit$] 
{\sl Inclusive scattering of polarized electrons (off unpolarized 
targets)}

From the difference between the ${\vec e}$--nucleus cross sections, 
with polarization of the electron parallel and antiparallel to the 
incident beam it is possible to measure the {\it asymmetry}:
\bea
{\cal A} &=& 
\frac{d^2\sigma^+ - d^2\sigma^-}{d^2\sigma^+ + d^2\sigma^-}
\nonumber\\
&\equiv&
{\cal A}_o\frac{ v_L R_{AV}^L(q,\omega) +
v_T R_{AV}^T(q,\omega) +v_{T'} R_{VA}^{T'}(q,\omega)}
{v_L R_L(q,\omega) + v_T R_T(q,\omega)}
\label{asymm}
\eea
which contains, in the numerator, the parity violating (PV) nuclear
response functions, stemming from the interference between the nuclear
electromagnetic and  weak (vector and axial) neutral currents. 
In the above
\beq
{\cal A}_o =\frac{G_F Q^2}{2\pi\alpha\sqrt{2}}
\approx 3.1\times 10^{-4}\tau \qquad\quad
\left(\tau=\frac{Q^2}{4M^2}\right)
\eeq
where $G_F$ is the Fermi constant, $\alpha$ the electromagnetic coupling 
constant, $Q^2=-q_{\mu}^2 >0$ and $v_{T'}=\sqrt{\tan^2\left(\theta/2\right) 
- (q_{\mu}^2/q^2)\tan(\theta/2)}$. It is worth mentioning that the PV 
response functions can be sensitive to novel aspects of the nuclear 
dynamics, different from the ones explored by the electromagnetic vertex.
In particular they give access to the isoscalar NN force and to the
neutrons' distribution in nuclei.

\item[$\diamondsuit$]
{\sl Exclusive versus inclusive electron scattering}

One--exclusive processes, like $(e,e',N)$, in which one nucleon is 
observed in coincidence with the outgoing electron, are more sensitive 
tests of the nuclear models. Indeed the probability of a 
nucleon to be emitted strongly reflects, for example, the momentum 
distribution inside the nucleus; striking differences are obviously 
found by employing models like a realistic shell model, the so--called 
Hybrid model or even the Fermi Gas model (the latter being, manifestly, 
not appropriate to deal with the details of nuclear structure).

A  variety of nuclear response functions can be separated by an 
adequate selection of the kinematics; in particular some interference 
between the longitudinal and transverse components of the electromagnetic 
current contribute to the one--exclusive cross sections, at variance 
with the inclusive ones. Thus one can test different components and 
matrix elements of the nuclear electromagnetic current, including 
delicate questions like the off--shellness of the nucleonic current 
in the nucleus, the contribution of meson exchange currents, the 
limitations of non--relativistic approaches.\cite{Amaro98,Cenni97}

Finally $(e,e',N)$ is fairly sensitive to the strong interaction of the 
ejected nucleon with the residual nucleus, the so--called final state 
interaction (FSI), which can be responsible for a sizeable distortion 
of the outgoing nucleon wave. Recently a semiclassical approach has 
been proposed\cite{Chanfray} to 
describe this specific aspect: it will be shortly 
mentioned here, since it provides  a natural bridge between nuclear 
matter and finite nuclei. The main research activity on the electromagnetic 
interaction in light nuclei is reported elsewhere\cite{Ciofi}.

\end{description}

Coming back to the main subject of this report, we shall now concentrate 
on the different theoretical methods which have been employed for an 
accurate evaluation of the electromagnetic response functions in 
nuclear matter. They can be roughly grouped into three categories:
\begin{enumerate}
\item 
The variational method, which is based on the Fermi Hypernetted 
Chain (FHNC) scheme
\item 
The functional method, which employs the path--integral approach
for developing a consistent bosonic loop expansion
\item 
The traditional perturbation theory, which encompasses the most 
frequently utilized models, from the Fermi gas (non--relativistic 
and relativistic), to the Hartree--Fock (HF) model combined together
with the important correlations provided by the Random Phase 
Approximation (RPA).
\end{enumerate}

In concluding this section we briefly mention the case of neutral
current (NC) 
neutrino-- and antineutrino--nucleus scattering: these processes
offer the possibility to extract information on the strange form factors
of the nucleon. For this purpose two quantities have been considered,
the $\nu-{\bar\nu}$  asymmetry\cite{Bai}:
\beq
{\cal A}_p = {\displaystyle
\frac{\left(\sigma\right)_{\nu p\rightarrow \nu p}^{NC} -
\left(\sigma\right)_{{\bar\nu}p\rightarrow {\bar\nu}p}^{NC} }
{\left(\sigma\right)_{\nu n\rightarrow \mu^- p}^{CC} -
\left(\sigma\right)_{{\bar\nu} p\rightarrow \mu^+ n}^{CC} }
}\, 
\label{nuasym}
\eeq
and the ratio of the cross sections for inelastic 
NC scattering of $\nu({\bar\nu})$ on nuclei, with the emission of a 
proton or, respectively, of a neutron:
\beq
{\cal R}^{\nu(\bar\nu)}_{p/n}=
\frac{\displaystyle\left( \sigma \right)^{NC}_{\nu({\bar\nu}),p} }
{\displaystyle \left(\sigma\right)^{NC}_{\nu({\bar\nu}),n} }
\, .
\label{ratio}
\eeq

Calculations of both (\ref{nuasym}) and (\ref{ratio}) have been 
performed, in an energy range from 200 MeV to 1 GeV, within two 
relativistic independent particle models (Fermi gas and shell model); 
the final state interactions of the ejected nucleon have been taken 
into account through relativistic Optical model 
potentials.\cite{Sam1,Sam2,Sam3}.
 
While the values of the cross sections significantly  depend on the 
nuclear model (especially in the lower energy range),
 the NC/CC neutrino--antineutrino asymmetry and the ratio p/n
 show a rather mild  dependence on the model and allow one to disentangle 
different values of the strangeness parameters entering into the weak NC 
axial and vector strange form factors of the nucleon. 

We remind that the $Q^2=0$ limit of the axial strange form 
factor, $g_A^s$, is closely related to the problem of the proton spin, 
a long lasting puzzle raised by the measurements, in the deep inelastic
polarized lepton scattering, of the polarized structure function $g_1$
of the proton. Moreover the information on the strange form factors 
of the nucleon extracted fron neutrino scattering is complementary 
to the similar information, which can be obtained in the above quoted 
PV polarized electron scattering experiments.

\section{ Electromagnetic response functions}

\subsection{Variational method}

The inclusive {\it transverse response function} of symmetric 
nuclear matter at saturation density, $R_T(q,\omega)$, has been studied 
by Fabrocini\cite{Fab1} within the Correlated Basis Function (CBF) 
perturbation theory. The main goal of this work is to ascertain how 
$R_T$ is affected by the NN correlations, with a special emphasis 
on the MEC contributions. 

CBF calculations are based upon a set of correlated wave functions
\beq
|n>={\cal S}\left[\prod_{i,j} f(i,j)\right]|n>_{FG}
\label{CBFstate}
\eeq
obtained by applying a symmetrized product of two--body correlation 
operators, $f(i,j)$, to the FG states $|n>_{FG}$. The following 
effective structure is adopted for $f$:
\bea
&&f(i,j)=\sum_{q=1,6} f^{(q)}(r_{ij}) O_{ij}^{(q)},
\label{fij}\\
O_{ij}^{(q=1,6)}&&= (1,{\sigma}_i\cdot{\sigma}_j, S_{ij})
\otimes(1,{\tau}_i\cdot{\tau}_j)
\label{operij}
\eea
$S_{ij}$ being the tensor operator; $f(i,j)$ depends upon a set of 
variational parameters, which are fixed by minimizing the expectation 
value of a realistic, non--relativistic Hamiltonian on the correlated 
ground state. The g.s. energy is calculated via the FHNC cluster 
summation technique, using the correlation corresponding to the 
Argonne $V_{14}$~+~Urbana~VII three--nucleon interaction model of 
ref.\cite{Adel21}.

The Jastrow (scalar) component of (\ref{operij}) accounts for the 
short range NN repulsion, while, among the remaining correlators, the
most relevant are the spin--isospin  and the tensor--isospin ones, 
which  mostly stem from  the one--pion exchange (OPE) long range part 
of the potential.

Configurations up to correlated one particle--one hole (1p--1h) 
intermediate states
are considered; the spreading due to the decay of particle (hole) 
states into 2p--1h (2h--1p) states is taken into account via a 
realistic optical potential model. 

The transverse response is given by
\beq
R_T(q,\omega) =\frac{1}{A}\sum_n \left\vert<0|{\vec j}({\vec q})|n>
\right\vert^2\delta(\omega-\omega_n)
\label{RTadel}
\eeq
where the sum goes over the intermediate excited states  $|n>$, with 
excitation energy $\omega_n$; ${\vec j}({\vec q})$ is the 
electromagnetic current operator 
\beq
{\vec j}({\vec q}) = {\vec j}^{(1)}({\vec q}) + {\vec j}^{(2)}({\vec q})
\label{curr}
\eeq
sum of the one--body, magnetic current [$ {\vec j}^{(1)}({\vec q}$] and
the two--body exchange currents [${\vec j}^{(2)}({\vec q})$]. For the 
latter, the Schiavilla--Pandharipande--Riska model\cite{SPR} is adopted,
which satisfies, by construction, the continuity equation with realistic 
$V_{14}$ Argonne and Urbana potentials. Currents due to intermediate 
$\Delta$--isobar excitations are also included. The sharp energy 
boundaries of the 1p--1h FG response are smoothed out by an appropriate
folding with a width $W(\omega)$:
\beq
R_T(q,\omega) =\frac{1}{\pi}\int\,d\omega'\, R_T^{1p-1h}(q,\omega')
\frac{W(\omega')}{(\omega-\omega')^2+[W(\omega')]^2},
\label{RTwidt}
\eeq
where $W(\omega)={\mathrm Im} W_o(\omega)/M^*$, $W_o$  being 
the optical potential and $M^*$ the nucleon effective mass.

%%%%%%%%%%%%%%%%%%%%   Figure 1 %%%%%%%%%%%%%%%%%%%%%%%%%%%%%%%%%%%%%%

\begin{figure}
%\vspace{8 cm}
\centerline{
\epsfxsize=11cm
\epsfysize=17.5cm \epsfbox{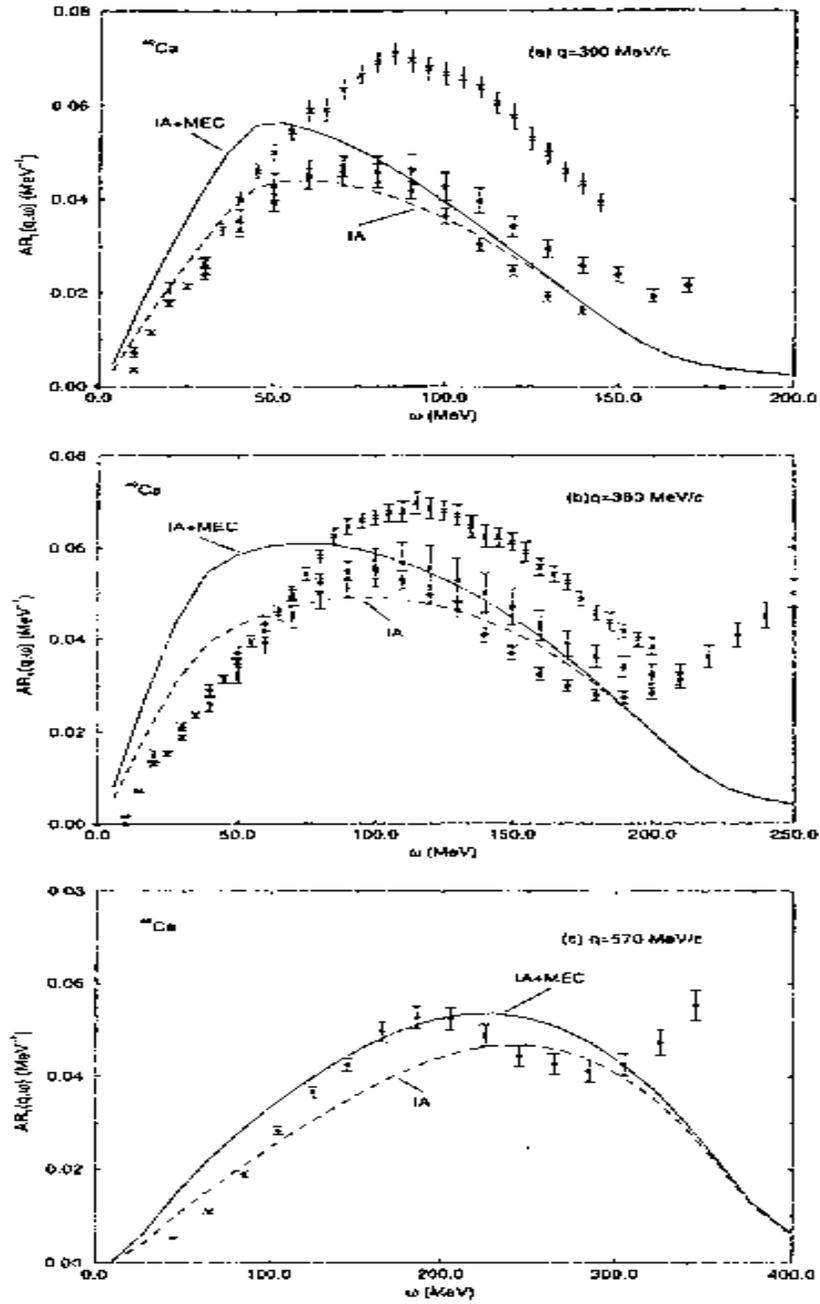}  
}
\caption[ ]{ Transverse response at $q=300$ (a), 380 (b), and 570 (c)
MeV/c for $^{40}$Ca and nuclear matter. See text.  }
\label{fig1}
\end{figure}

Fig.~1 shows the comparison of the nuclear matter response evaluated in
ref.\cite{Fab1} and the experimental data on $^{40}$Ca; both the impulse
approximation (IA) and the full calculation (MEC+IA) are shown. The data 
are taken from from ref.\cite{Adel4} ($\times$), ref.\cite{Adel6} (circles)
and ref.\cite{Adel9} (black circles).
The global contribution of the two--body currents turns out to be 
positive and provides an enhancement of the one--body response, 
ranging from $\sim 20\%$ for the lower momenta (300~MeV/c) to $\sim 10\%$
for the higher ones (about 500 MeV/c). The tensor--isospin component
of the correlation is crucial to obtain these results.

Several variational calculations have been performed in the past for the 
{\it longitudinal response function} as well, both in nuclear matter and
light nuclei. In a recent work Amaro {\it et al.}\cite{Fab2} have 
tested a simple approximation to deal  with the short range correlations 
(SRC) affecting the 1p--1h 
excited states, which build up the nuclear response functions.

The basic idea of the model (previously tested for the  ground state 
properties)  consists in truncating the CBF expansion in 
such a way to retain only those terms containing a single Jastrow--type
correlation line,
\beq
h(r)=f^2(r)-1\, .
\label{Jastrow}
\eeq
The expansion in powers of $h(r)$ has the property of conserving 
the proper normalization of the correlated many--body wave function.
The nuclear charge is conserved as well, at variance with other  
truncation schemes. The correlation employed is the scalar component 
of a complicated state dependent correlation, fixed to minimize the 
nuclear binding energy in a FHNC calculation with the Urbana $V_{14}$
NN potential.

%%%%%%%%%%%%%%%%%%%%   Figure 2 %%%%%%%%%%%%%%%%%%%%%%%%%%%%%%%%%%%%%%

\begin{figure}
%\vspace{8 cm}
\centerline{
\epsfxsize=11cm
\epsfysize=11cm \epsfbox{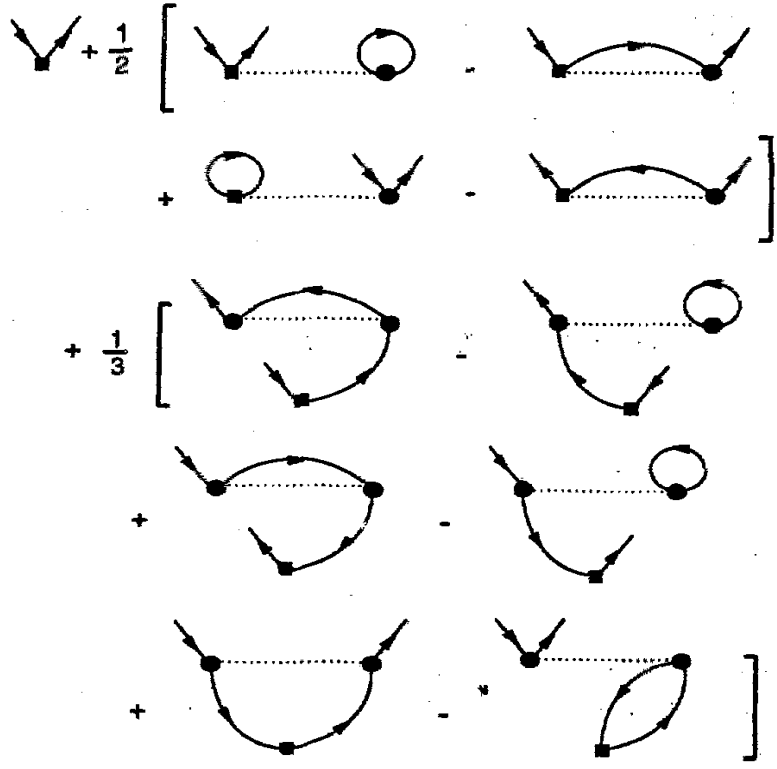}  
}
\caption[ ]{Diagrams considered in the model of ref.\cite{Fab2}. The dotted
lines represent the correlation function. The oriented lines represent 
particle and hole wave functions. The black circle indicates an integration
point, while the black squares indicates the integration point, where 
the charge operator is acting.   }
\label{fig2}
\end{figure}

Fig.~2 shows the diagrams retained in \cite{Fab2}. The 
comparison between the calculation of the longitudinal response 
function in the full FHNC and in the proposed approximation shows 
that the difference between the corresponding results does not exceed a 
few parts $\times 10^{-4}$, in a fairly large interval of momentum
transfers. This proves the reliability of the truncated scheme also 
for dealing with 1p--1h excited states; it needs to be further tested,
however, for the 2p--2h responses.

\subsection{Functional (path integral) method}

The path integral approach offers a consistent theoretical foundation
of the nuclear response functions. Indeed it employs the approximation 
schemes of QFT, namely the perturbative approach and the loop 
expansion, to provide a fully consistent criterium of selecting 
classes of diagrams, still preserving the fundamental symmetries and
invariances of the choosen model Lagrangian. 

Starting from a Lagrangian which describes interacting nucleons, 
pions and eventually $\rho$ mesons, all coupled to an external 
electromagnetic field, one derives the nuclear 
 polarization propagator and response functions through
derivatives of the generating functional
\bea
Z[j_\mu^{ext}] &&= \frac{1}{{\cal N}}\int {\cal D}\left[ {\bar\psi},
\psi,\Phi,A_\mu\right] \exp\left\{i\int dx[j_\mu^{ext}A_\mu]\right\}
\nonumber\\
&&\times\exp\left\{i\int dx\left({\cal L}+{\cal L}_{em}+J_\mu A^\mu
+B_{\mu\nu} A^\mu A^\nu\right)\right\}
\label{funZ}
\eea
where $\psi,{\bar\psi}$ are the nucleon fields, $\Phi$ and $A_{\mu}$,
respectively, the pion and electromagnetic fields and 
$\j_{\mu}^{ext}$ the external source of the latter. In the above
\beq
{\cal L}= {\bar\psi}(i\gamma^\mu\partial_\mu-M)\psi +
{\scriptstyle\frac{1}{2}}(\partial_\mu \Phi)^2 -\frac{m_\pi^2}{2}\Phi^2
-ig{\bar\psi}\gamma_5\tau\psi\cdot\Phi
\label{funLag}
\eeq
is the model Lagrangian (without the $\rho$ meson) and
\bea
J_\mu &=& e{\bar\psi}{\scriptstyle\frac{1}{2}}(1+\tau_3)\gamma_\mu\psi +
e[\Phi\times\partial_\mu\Phi]_3
\label{funcurrent}\\
B_{\mu\nu} &=& e^2g_{\mu\nu}\Phi^+\Phi^-
\label{fun2pion}
\eea
the required couplings of nucleons and pions to the e.m. field.
The results obtained, within this formalism, for the longitudinal and 
transverse responses in the inclusive electron scattering are shown in 
the work by Cenni {\it et al.}\cite{Cenni1}

It is worth mentioning that an important connection between the 
variational technique and the path integral approach has been recently 
explored in ref.\cite{Cenni2}. These authors suggest a new method, based on
the variation, within a path integral framework, of a trial Hamiltonian: 
it turns out that a particular choice of the latter
corresponds exactly to the use of a Jastrow correlated Ansatz for the 
wave function in the FHNC approach. Thus the new formalism generalizes 
the FHNC and CBF techniques, allowing for their extension to 
relativistic and field theoretical problems.

\subsection{Perturbation theory}

Several approaches based on perturbative schemes have been applied to 
the evaluation of nuclear response functions both for inclusive 
(parity conserving and parity violating) and exclusive processes. Most 
calculations are based on non--relativistic Hamiltonians and currents, 
but attempts to include relativistic effects, especially in the 
electromagnetic currents and vertices, have been done.

Many--body perturbative techniques have been  developed to 
treat quasielastic electron scattering in nuclear matter. 
The Green's function method is particularly suited to express 
inelastic cross sections through the imaginary part of the polarization 
(particle--hole) propagator; nuclear matter results compare fairly 
well with the experimental data for energy transfers in the region of 
the quasielastic peak and above, providing the $\Delta_{33}$ resonance 
and mesonic degrees of freedom are  incorporated in the 
 nuclear matter polarization propagators. 

The role played by the NN interaction in reshaping the nuclear 
matter response functions with respect to the pure FG one has been widely 
investigated, by employing both phenomenological effective interactions
and G--matrix parameterizations based on the one--boson  exchange 
potential. Starting from the naive Fermi gas description, one can easily 
obtain the Hartree--Fock (HF) or Brueckner--Hartree-Fock (BHF) 
polarization propagators; the next step is then to include RPA 
correlations, both in the ``minimal'' version of the so--called 
{\sl ring} diagrams and in the framework of the fully antisymmetrized 
RPA. 

Numerical calculations of the HF+RPA response functions can require 
large computing time. Indeed, although in nuclear matter the 
ring approximation for the polarization propagator 
can be analytically evaluated, as soon as 
one dresses the particle and hole propagators with HF self--energies, 
even the simple HF p--h propagator, $\Pi_{HF}(q,\omega)$, requires 
numerical integrations when the HF self--energy is derived from any 
realistic NN interaction.

Moreover the simple algebraic equation for the ring polarization 
propagator turns into an infinite series of complex integrals when
the exchange matrix elements of the p--h interaction are taken into 
account in the fully antisymmetrized RPA, which is diagrammatically 
shown in Fig.~3. An approximate, but reliable, treatment of 
these contributions is thus compulsory. 

%%%%%%%%%%%%%%%%%%%%   Figure 3    %%%%%%%%%%%%%%%%%%%%%%%%%%%%%%%%%%%%%%

\begin{figure}
\centerline{
\epsfxsize=11cm
\epsfysize=5cm \epsfbox{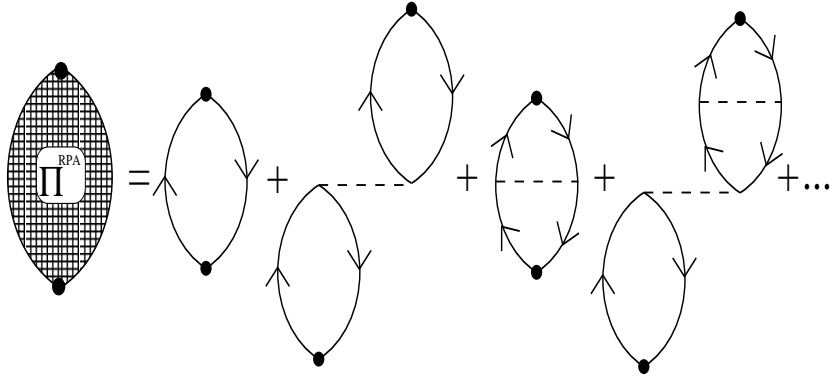}  
}
\caption[ ]{Diagrammatic representation of the perturbative expansion 
for the polarization propagator in RPA. The dashed line represents the
particle--hole interaction $V$. }
\label{fig3}
\end{figure}

This has been achieved with the method of the {\sl continued fraction} 
(CF) expansion\cite{Arturo}, which will be briefly sketched in the 
following. The CF--like expansion for the polarization propagator can
be written as 
\beq
\Pi^{RPA}={\displaystyle\frac{\Pi^{(0)}}
{\displaystyle 1-A-\frac{\displaystyle B}{\displaystyle 1-C-
\frac{D}{1-\dots}}}}
\label{Picf}
\eeq
where $\Pi^{(0)}$ is the free (or HF) particle--hole propagator and
all symbols appearing in (\ref{Picf}) are functions of both the energy 
and momentum transfers. The CF approach at n--th order reproduces 
exactly the perturbative series expressing the antisymmetrized RPA 
at the same order while it approximates higher orders. 
Setting, for example, 
\beq
\Pi^{RPA} =\sum_{n=0}^{\infty}\, \Pi^{(n)}
\label{pisum}
\eeq
the first order in CF provides the following approximation for 
$\Pi^{(n)}$:
\beq
\Pi^{(n)}\simeq \Pi^{(0)}\left[\frac{\Pi^{(1)}}{\Pi^{(0)}}
\right]^n,
\eeq
where $\Pi^{(1)}\equiv \Pi^{(0)}4V\Pi^{(0)} +\Pi^{(1)ex}$ is the sum 
of the direct and exchange first order terms. With this approximation 
the summation in (\ref{pisum}) is trivial, yielding
\beq
\Pi^{RPA}_{CF1} =
\frac{\Pi^{(0)}}{1- \Pi^{(1)}/\Pi^{(0)}}
=\frac{\Pi^{(0)}}{1- 4V\Pi^{(0)}- \Pi^{(1)ex}/\Pi^{(0)}}\, .
\label{Picf1}
\eeq
The next step extends the approximation to the exact second order 
term; the corresponding RPA propagator reads:
\bea
\Pi^{RPA}_{CF2} &=&
\frac{\Pi^{(0)}} {1- \Pi^{(1)}/\Pi^{(0)} -
\{ \Pi^{(2)}/\Pi^{(0)} - [\Pi^{(1)}/\Pi^{(0)}]^2 \} }
\label{Picf2}\\
&=&\frac{\Pi^{(0)}}{1- 4V\Pi^{(0)}- \Pi^{(1)ex}/\Pi^{(0)} 
- \{ \Pi^{(2)ex}/\Pi^{(0)} - [\Pi^{(1)ex}/\Pi^{(0)}]^2 \} }
\, ,
\nonumber
\eea
where the approximation provided by the first order CF expansion
has  been subtracted off. 

Numerical calculations for the longitudinal and transverse responses, 
using a G--matrix based on the Bonn potential, show that the CF
expansion rapidly converges, giving a few percent correction between
the first and the second order CF results. On the contrary, 
the differences between ring and RPA (in the CF1 or CF2 approximation) 
responses appear to be sizeable, both in the longitudinal and in the 
transverse channel, with, perhaps,  
the exception of the spin--isovector channel, as it is illustrated in 
Fig.~3. 
We also remind here that a very useful (and accurate) approximation,
consisting in a ``bi--parabolic'' parameterization of the HF self--energy 
of the nucleon\cite{Bai96}, allows to express analytically $\Pi_{HF}$, thus
considerably reducing the computational time.

%%%%%%%%%%%%%%%%%%%%   Figure 4    %%%%%%%%%%%%%%%%%%%%%%%%%%%%%%%%%%%%%%

\begin{figure}
\centerline{
\epsfxsize=11cm
\epsfysize=11cm \epsfbox{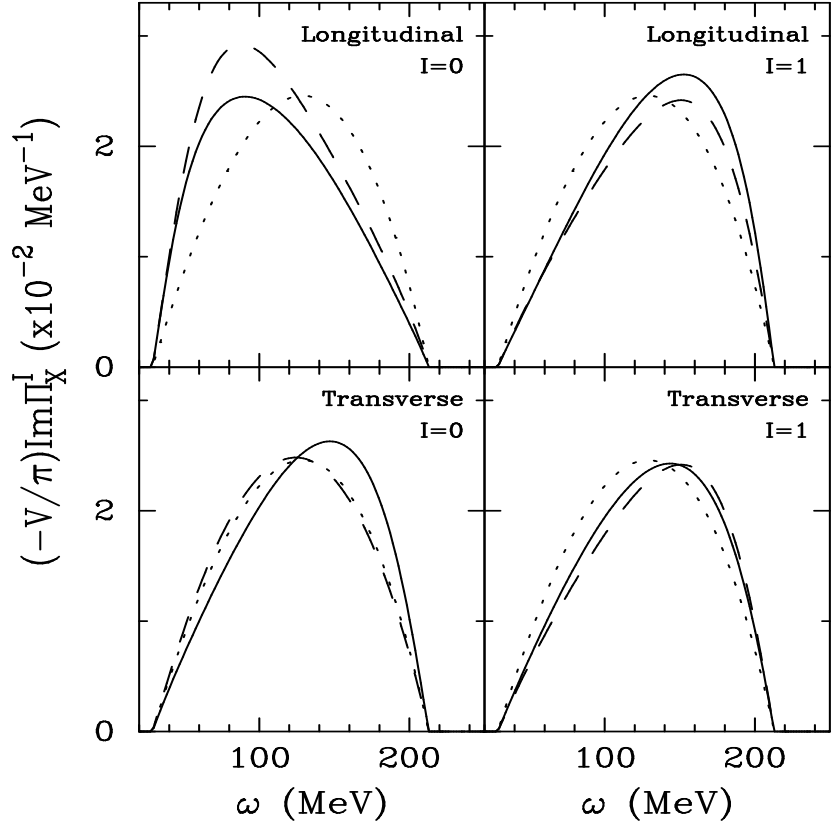}  
}
\caption[ ]{Nuclear matter responses for $k_F=195$~MeV/c at $q=500$~MeV/c:
the dotted line corresponds to the free response, the dashed one to the
ring approximation and the solid line to the RPA--CF1 response. The
kinematics is relativistic and the particle--hole interaction is derived 
from a realistic G--matrix. }
\label{fig4}
\end{figure}

Within the perturbative approach it is also possible to deal with 
{\it relativistic effects}, which might become important for energy/momentum 
transfers larger than 0.5~GeV or so. Besides the corrections stemming from 
the relativistic kinematics and vertices, which have been completely 
taken into account  in the framework of the Fermi gas model, 
several investigations, based on a fully 
relativistic Lagrangian with nucleonic and pionic degrees of freedom,
both coupled to the external electromagnetic field, have been performed.
These approaches 
usually include both the single--nucleon e.m. current and the two--body
MEC and account for relativistic effects within an expansion in terms
of suitably ``small'' variables, like the energy/momentum transfers in
units of twice the nucleon mass ($\kappa=q/2M, \lambda=\omega/2M$).
A new relativistic expansion for the matrix elements of one-- and 
two--body e.m. currents has been recently proposed by Amaro 
{\it et al.}\cite{Amaro98} in terms of the parameters $\eta_i\equiv p_i/M$ 
(characteristically of the order of 1/4),
$\{p_i\}$ being  the initial--state nucleon momentum inside the nucleus: 
clearly this formalism will be especially suited in the regime of GeV 
energies, where, on the contrary, $\kappa$ and $\lambda$ are no longer small
with respect to 1.

\subsubsection{y--scaling and Coulomb sum rule}

\par
The concepts of y--scaling and Coulomb sum rule have been throughly
discussed in the past since they offer important tests for the 
accuracy and reliability of the nuclear model description 
underlying the calculation of the electromagnetic response functions.
Both of them require a careful investigation in order to 
disentangle, from the full expression of the longitudinal and 
transverse response functions, those dividing factors such that the
reduced responses have scaling properties or (for what concerns the
longitudinal response) satisfy the Coulomb sum rule. This task is
not trivial since, as shown by  the Relativistic Fermi Gas (RFG) 
model, the e.m. form factors to be divided by 
are not simply factorizable, due to the structure of the relativistic
one--nucleon current. 

In the work by Barbaro {\it et al.}\cite{Bai98} the dividing factors
$G_L,\,G_T$ are constructed in the RFG for the longitudinal and 
transverse response functions in such a way that the reduced responses,
so obtained, scale; another dividing factor, $H_L$, yields a (different)
reduced response which fulfills the Coulomb sum rule (CSR). The relevant 
point here is that $G_L,\,G_T$ and $H_L$ are all found to be only
weakly model--dependent, thus providing essentially universal 
dividing factors. 
It becomes thus possible to test whether specific models scale and
satisfy the CSR. This investigation has been carried out for 
 the Hybrid model\cite{Cenni97}  and the Quantum Hadrodynamic 
model\cite{Walecka} (QHM), each one with different 
types of interaction effects which go beyond the strict RFG.
It is found that while the Hybrid model scales and obeys the CSR,
the QHM does neither. The Coulomb and higher order sum rules have
also been investigated by Amore {\it et al.}\cite{Amore} in the RFG,
by employing two different methods, namely by exploiting the scaling
properties of the longitudinal response function  and by enforcing
the completeness of the states in the space--like domain via the
Foldy--Wouthuysen transformation.

\subsection{Exclusive versus inclusive electron scattering scattering}

The connections between exclusive and inclusive electron--nucleus
scattering within the framework of the plane--wave impulse
approximation (PWIA) have been investigated in a work by Cenni {\it
et al.}\cite{Cenni97}. These authors test the interplay between the (model
independent) kinematical constraint and the (model
dependent) features of the spectral
function in providing the exclusive (and inclusive) nuclear responses.
The RFG and the Hybrid model are employed to provide a link between
finite and infinite Fermi systems and to assess the impact of the 
confinement of the struck nucleons on the inclusive charge response.
One of the important outcomes of this work is that, with an energy--shift
and rescaled Fermi momentum, an effective RFG response can be obtained
whose first three energy--weighted moments agree quite well with the
analogous quantities evaluated within the (confined) Hybrid model.

A more recent work\cite{Chanfray} evaluates  semi--classically 
the  exclusive $(e,e',p)$ cross sections, yet partially preserving
the simplicity of nuclear matter calculation, but going beyond the
PWIA. Indeed this approach allows to include the distortion of the
outgoing nucleon wave (the so--called final state interaction, FSI)
at least within the mean field approximation.

%%%%%%%%%%%%%%%%%%  FIG. 5 %%%%%%%%%%%%%%%%%%%%%%%%%%%%%%%%%

\begin{figure}
\centerline{
\epsfxsize=11cm
\epsfysize=5.5cm \epsfbox{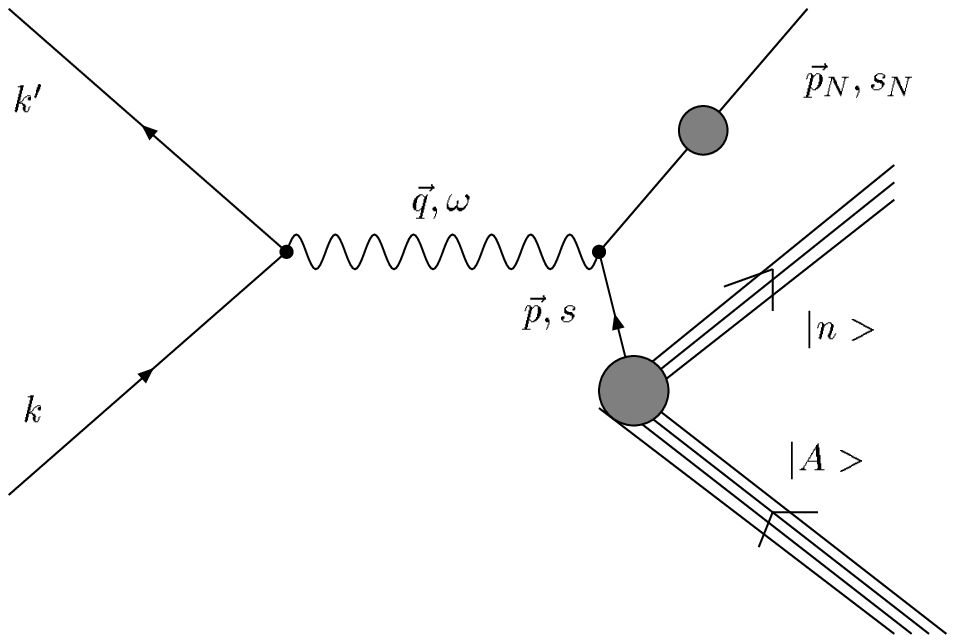}  
}
\caption[ ]{   }
\label{fig5}
\end{figure}

Sticking to the IA, the cross section for the $(e,e',p)$ scattering
process, which is schematically illustrated in Fig.~5, reads:
\beq
\frac{d^4\sigma}{d\epsilon' d\Omega_e dE_\N d\Omega_\N}
= E_\N p_\N \left(\frac{d\sigma}{d\Omega_e}\right)_{e\N}
 S({\vec p}_m,{\cal E}).
\label{exclcr}
\eeq
In the above:
\beq
\left(\frac{d\sigma}{d\Omega_e}\right)_{e\N}=\sigma_{Mott}
\frac{1}{\cos^2(\theta/2)}\frac{m_e^2}{\epsilon\epsilon'}
\eta^{\mu\nu}\frac{M^2}{E({\vec p})E_\N}
W_{\mu\nu}({\vec p},{\vec p}+{\vec q})
\label{sncr}
\eeq
is the single nucleon cross section, while
\beq
 S({\vec q},q_0)=\frac{1} {(2\pi)^3} \frac{M} {E({\vec q})}
\frac{<A|{\hat a}^{\dagger}_q 
\delta\left( q_0 - {\hat{\cal H}}-\mu\right){\hat a}_q|A>}{<A|A>}
\label{spectral}
\eeq
is the nuclear spectral function, which in (\ref{exclcr}) depends upon
the missing momentum ${\vec p}_m={\vec q}-{\vec p}_\N$ and the missing
energy ${\cal E}=E_\N-\omega$ (see Fig.~5 for the other symbols).

Different nuclear models generally provide different spectral functions,
as one can see, for example, by comparing the ones obtained in the RFG,
\beq
S^{RFG}({\vec p},{\cal E})=\frac{\Omega} {(2\pi)^3}
\theta(k_F-p)\delta\left\{{\cal E}-\sqrt{k_F^2+M^2}-
\sqrt{p^2+M^2}\right\}
\label{sfRFG}
\eeq
and in the Hybrid model (harmonic oscillator bound states combined
with a continuum of unbound states):
\beq
S^{HM}({\vec p},{\cal E})=\sum_{N=0}^{N_{max}}
\delta\left\{ {\cal E}-(N_{max}-N)\omega_0\right\} n_{N}(p),
\label{sfHM}
\eeq
where
\beq
n_{N}(p)=\sum_{n,\ell=N-2n}\frac{2\ell+1}{4\pi}
|\varphi_{n\ell}(p)|^2
\eeq
is the momentum distribution for the N--th shell. The spectral
function (\ref{sfHM}) is illustrated in Fig.~6, where a dashed line
is drawn in correspondence with the support for the spectral function
of the RFG, eq.~(\ref{sfRFG}), which is infinite along that line and
zero elsewhere.

%%%%%%%%%%%%%%%%%%%%%%%  FIG. 6   %%%%%%%%%%%%%%%%%%%%%%%%%%%%%%%

\begin{figure}
\centerline{
\epsfxsize=11cm
\epsfysize=9cm \epsfbox{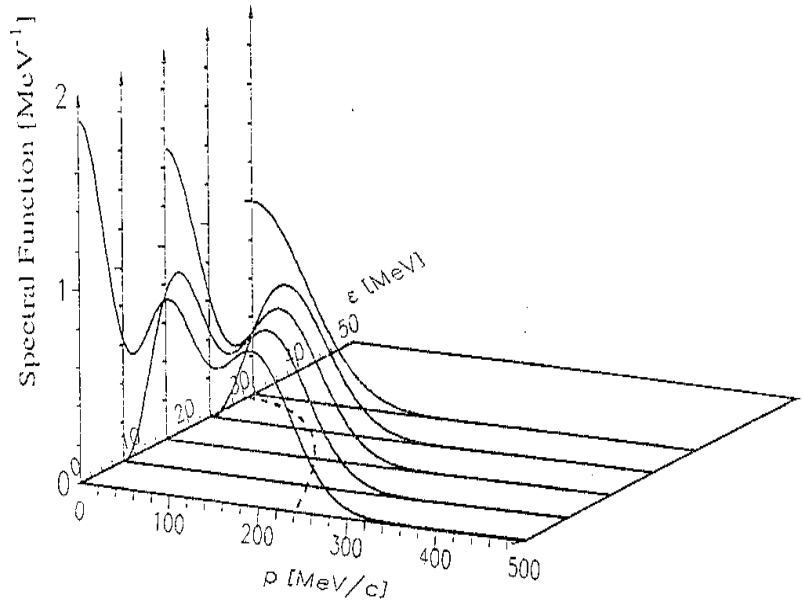}  
}
\caption[ ]{The spectral function in the Hybrid Model. }
\label{fig6}
\end{figure}

Within the semiclassical framework, the Wigner transform of the
spectral function reads:
\beq
\left[ S({\cal E})\right]_W(R,{\vec p}) =
\theta\left[\epsilon_F -\frac{p^2}{2M^*}-V(R)\right]
\delta\left\{ {\cal E}-\left[\epsilon_F-\frac{p^2}{2M^*}-V(R)
\right]\right\}
\label{sfsemicl}
\eeq
where $V(R)$ is a suitable shell model potential (e.g. the harmonic
oscillator or the Woods--Saxon potentials) and $\epsilon_F\equiv
k_F^2(R)/2M^* +V(R)$ the local Fermi energy.
It is shown in ref.\cite{Chanfray} that in the semiclassical approach
it is possible  to  account for the distortion of the ejected nucleon
(illustrated by the black dot in Fig.5) in a relatively simple way:
an appropriate distortion operator is introduced, for which 
two heuristic assumptions are made,  corresponding to the eikonal 
and to the uniform approximations, respectively. 
The corresponding exclusive cross sections turn out to
be quite different, hence stressing the relevance of the FSI in the
analysis of the exclusive processes.

%%%%%%%%%%%%%%%%%%%%%%%%%%%%%%%%%%%%%%%%%%%%%%%%%%%%%%%%%%%%%%%%%%

\section{Strange probes and strange matter}

\subsection{$K^+$--nucleus scattering}

The quasielastic $K^+$--nucleus scattering has been thoroughly
investigated by De Pace {\it et al.}\cite{Art1,Art2,Art3}.
This subject shares several motivations with the items in the
previous sections. The first one relies on the relatively small
(as compared with strong interaction processes) 
$K^+ N$ cross sections: as a consequence the $K^+$ projectiles
can penetrate deeply inside the nucleus, similarly to electrons
and photons, thus allowing to investigate collective effects in
the nuclear response. The second reason of interest is the
possibility of exploring scalar--isoscalar excited states, which
are the dominant channel for the $K^+$--nucleon coupling. This
channel can be partly explored in the longitudinal electron scattering
response functions, but an isospin separation in addition to the
customary longitudinal--transverse one has never been, till now,
carried out in $(e,e',X)$ experiments.\footnote{Such a separation 
should become possible with PV electron scattering experiments and 
it will be of extreme interest to compare the latter with the $K^+$ 
quasielastic scattering.}
Finally the observed excess in elastic $K^+$--nucleus scattering
cross sections seems to require, to be explained, some enhancement
induced by the nuclear medium itself.

A simple phenomenological approach, which expresses the above
mentioned cross sections by means of an effective number of
participant ($N_{eff}$),
\beq
\frac{d^2\sigma}{d\Omega d\omega} = N_{eff}
\frac{d\sigma_{K^+N} }{d\Omega} R(q,\omega),
\label{Neff}
\eeq
seems to require values of $N_{eff}$ too large with respect to
what expected in the Glauber theory; moreover (\ref{Neff}) fails
in describing the observed collective phenomena at low momentum
transfer.

The model proposed in  \cite{Art1} allows to evaluate the nuclear
response $R(q,\omega)$ within a continuum RPA, using an effective
G--matrix particle--hole interaction and a Woods Saxon mean field,
complemented by an appropriate spreading width of the ph states.
The kinematics is relativistic. An improved Glauber theory is used
for the reaction mechanism, including one-- and two--step
contributions.

%%%%%%%%%%%%%%%%%%%%%%%  Fig. 7  %%%%%%%%%%%%%%%%%%%%%%%%%%%%%%%%%%%%

\begin{figure}
\centerline{
\epsfxsize=12cm
\epsfysize=8cm \epsfbox{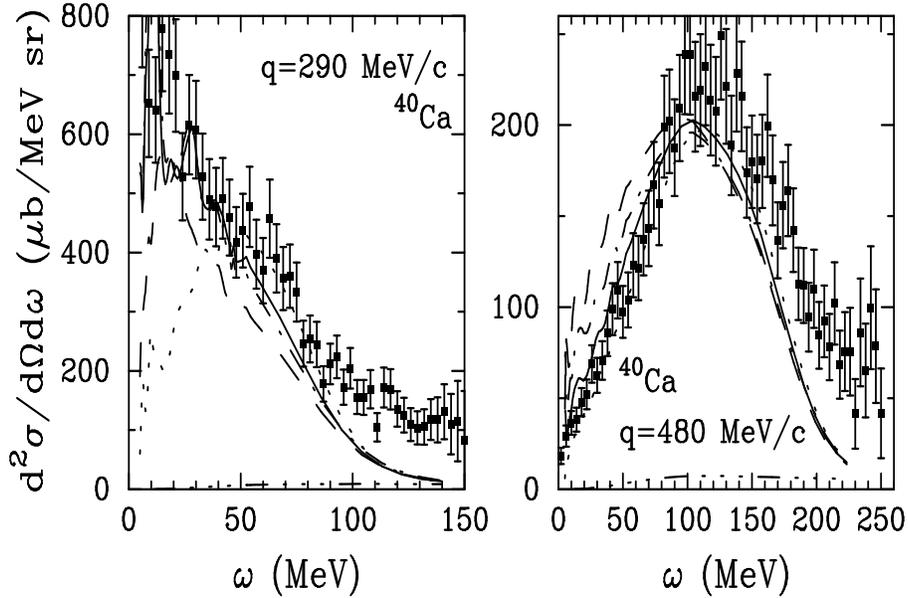}  
}
\caption[ ]{The $K^+ - ^{40}$Ca cross sections at $q=290$~MeV/c
(left) and $q=480$~MeV/c (right); data are taken from ref.\cite{Korman}.
For the explanation of the theoretical curves see text. }
\label{fig7}
\end{figure}

In Fig.~7 we show an example of the results obtained in this
framework: the experimental data for $K^+-^{40}$Ca scattering are
compared with the 1 step (free)+ 2 step calculation (dotted line),
the 1 step (using the $N_{eff}$ formula)+ 2 step (dot-dashed line)
and the 1 step (RPA)+ 2 step (dashed line). While the free response
is clearly inadequate to reproduce the data, the other two calculations
fail in reproducing the excess cross section appearing in the low
energy region at small momentum transfers. However if one empirically
reduces by about 50\% the effective ph interaction in the scalar
isoscalar channel, then the 1 step (RPA)+ 2 step turns out to be in
very good agreement with the experimental data, in the whole range
of momentum transfers and nuclei ($^{12}$C and $^{40}$Ca) explored.
This is a new, challenging information on the ph force acting 
in this channel.

\subsection{Hypernuclei (structure and decay)}

In the previous paragraph we have considered the interaction of a
strange particle, the $K^+({\bar s}u)$, with the nuclear medium.
Should a $K^-(s{\bar u})$ hit a nucleus, most of the times
it gets absorbed by a nucleon, leading to the formation of a
hypernucleus, namely a nucleus containing one (or more) hyperon $(sqq)$.
The lightest and more easily produced hyperon is the $\Lambda$ particle
and a great interest is focussed on the investigation of the ground
state properties and decay mechanisms of $\Lambda$--hypernuclei.
Several experimental data are already
available on light and medium hypernuclei,
but a reacher set of measurements are expected from the FINUDA experiment,
which is just starting operating in the Frascati National Laboratory of INFN.

The {\bf weak decay width} of $\Lambda$--hypernuclei can be evaluated with
the Green's function (polarization propagator) method, which is
conveniently implemented in nuclear matter and extended to finite
systems through the so--called Local Density Approximation (LDA).
The width of the hypernucleus is derived from the $\Lambda$ self--energy
\beq
\Gamma_{\Lambda}(k,\rho)= -2{\mathrm Im}\Sigma_{\Lambda}(k,\rho)
\label{GammaL1}
\eeq
after integrating over the $\Lambda$ momentum distribution.

Three different mechanisms contribute to the weak decay of a $\Lambda$
in the nucleus:
\begin{itemize}
\item the {\it mesonic decay},
\[ \Lambda\longrightarrow \pi N \qquad (\Gamma_{M}), \]
which also occurs in free space and is relevant only in light hypernuclei,
since the Pauli blocking prevents the low--momentum outgoing nucleon to
be emitted in a medium--heavy nucleus;
\item the {\it two--body non--mesonic (NM) decay}
\[ \Lambda N \longrightarrow NN \qquad (\Gamma_1), \]
\item the {\it three--body non--mesonic decay}
\[ \Lambda NN \longrightarrow NNN \qquad (\Gamma_2). \]
\end{itemize}

%%%%%%%%%%%%%%%%%%%%   Figure 8    %%%%%%%%%%%%%%%%%%%%%%%%%%%%%%%%%%%%%%

\begin{figure}
\centerline{
\epsfxsize=11cm
\epsfysize=8cm \epsfbox{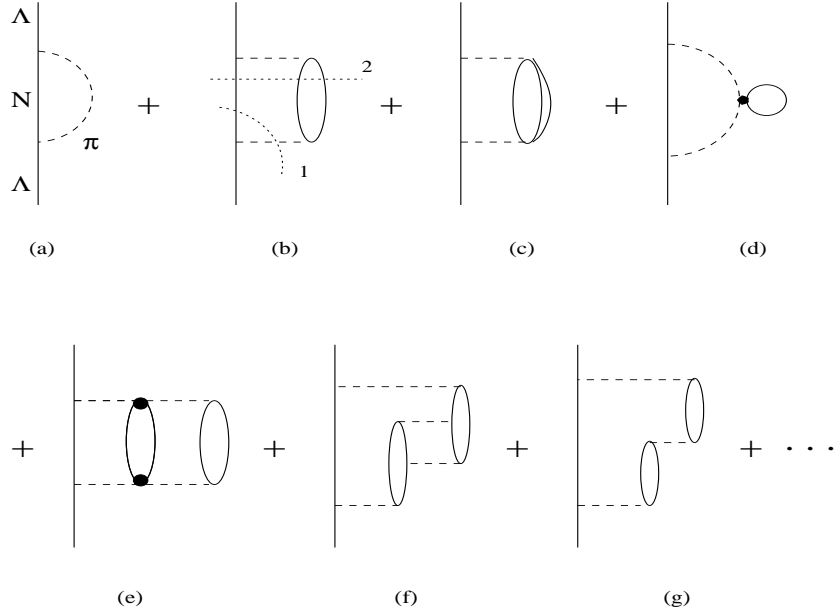}  
}
\caption[ ]{Lowest order terms for the $\Lambda$ self--energy in
nuclear matter. }
\label{fig8}
\end{figure}

A new evaluation of the $\Lambda$ self--energy has been carried 
out\cite{Garba}, which includes the 1p-1h and 2p-2h polarization 
propagators (related to $\Gamma_1$ and $\Gamma_2$, respectively) 
in a RPA scheme. Fig.~8 illustrates  a few examples 
of the diagrams whose imaginary part is contributing to the 
hypernuclear decay width.
The particle--hole interaction embodies the
exchange of $\pi$ and $\rho$ mesons, together with short range 
repulsive correlations described by the two Landau--Migdal 
parameters, $g'$ and $g'_{\Lambda}$ (the latter being used when 
one vertex of the exchanged pion coincides with the weak $\Lambda\pi N$
vertex). The nuclear matter calculation is adapted to finite nuclei
using the LDA, which amounts to replace the constant Fermi momentum
with a local one, expressed by $k_F(r)=[3\pi^2\rho(r)/2]^{1/3}$, 
$\rho(r)$ being the nuclear density. The width associated to the decay
of a $\Lambda$ with momentum $k$ is then
\beq
\Gamma_{\Lambda}({\vec k})=\int d{\vec r} |\psi_{\Lambda}({\vec r})|^2
\Gamma_{\Lambda}[{\vec k},\rho(r)]
\label{GammaL2}
\eeq
where $\psi_{\Lambda}({\vec r})$ is the $\Lambda$ wave function in the
nucleus. The results one obtains turn out to be
very sensitive to the latter, the best choice corresponding to
a Woods Saxon potential well which reproduces the measured $s$ and 
$p$ energy levels of the $\Lambda$ in the nucleus. 
Finally the total width of the hypernucleus is obtained by
integrating (\ref{GammaL2}) over the momentum distribution of the 
$\Lambda$ itself:
\beq
\Gamma_{\Lambda}=\int d{\vec k}\, |A_{\Lambda}({\vec k})|^2\, 
\Gamma_{\Lambda}({\vec k}).
\label{GammaL3}
\eeq
The results obtained within this framework are illustrated in Fig.~9,
where the decay width of  several hypernuclei (from the ``light'' $^{12}$C 
to the heaviest $^{209}$Bi and $^{238}$U) is shown, in units of the free
$\Lambda$ decay width, $\Gamma_{\mathrm free}$, as a function of the mass 
number. The separate contributions, $\Gamma_M$, $\Gamma_1$ and $\Gamma_2$,
are also displayed.

%%%%%%%%%%%%%%%%%%%%   Figure 9    %%%%%%%%%%%%%%%%%%%%%%%%%%%%%%%%%%%%%%

\begin{figure}
\centerline{
\epsfxsize=10cm
\epsfysize=10cm \epsfbox{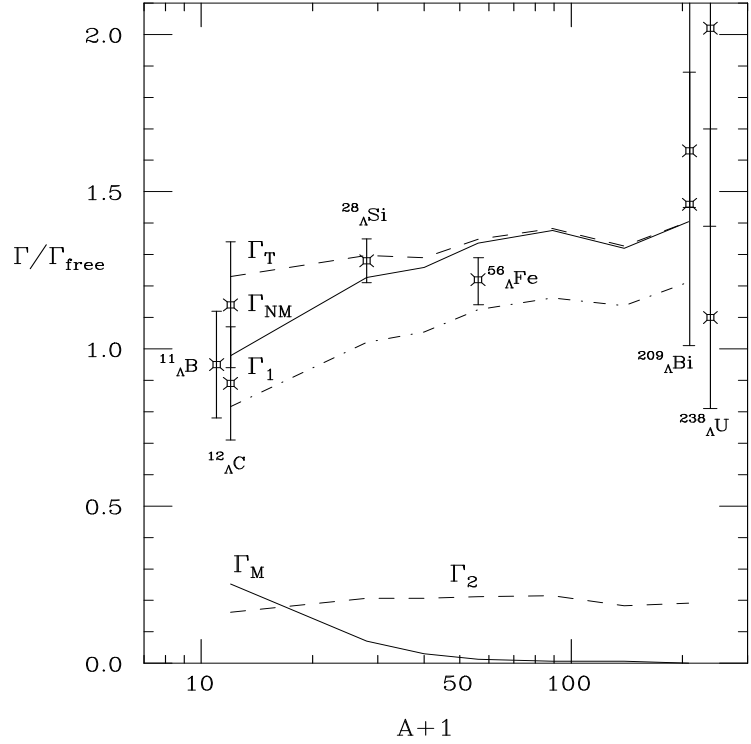}  
}
\caption[ ]{ $\Lambda$ decay widths in finite nuclei as a function of
the mass number $A$; the labels of the various curves correspond to:
the mesonic width ($\Gamma_M$), the one-- ($\Gamma_1$) and two--body
($\Gamma_2$) induced decay widths, the total non--mesonic width 
($\Gamma_{NM}=\Gamma_1+\Gamma_2$) and the total ($\Gamma_T$), sum 
of all independent contributions. }
\label{fig9}
\end{figure}

The theoretical results are in good agreement with the data over the
whole hypernuclear mass range explored. The saturation property of the
$\Lambda N\to NN$ interaction in nuclei clearly appears in the
flattening of the decay width as the mass number increases. In spite of
these satisfactory results, there remains to be explained the large
experimental value of the ratio $\Gamma_n/\Gamma_p$, which does not seem
to be in agreement with the present theoretical schemes.

Concerning the {\bf ground state properties of hypernuclei} we should
mention here a research project by Co {\it et al}\cite{Co99}, who intend
to investigate the structure of hypernuclei within the FHNC scheme.
Starting from a suitable basis of single particle wave functions
(Woods--Saxon and/or HF wave functions with finite range interaction)
a Jastrow correlated many--body state will be obtained by developing
a FHNC formalism for a system of A nucleons plus one impurity (the
hyperon). Binding energy and first excited states can also be calculated.

\section{Asymmetric nuclear matter}

\subsection{Equation of state of asymmetric matter}

Nuclear matter is usually considered as an isospin symmetric system,
$Z=N=A/2$, where the Coulomb interaction between protons is switched off.
In reality  heavier nuclei are more and more isospin asymmetric, the
larger number of neutrons being there to balance the increasing Coulomb
repulsion. An interesting, ideal extension of these systems is the
asymmetric nuclear matter, where the neutron excess is measured by the
asymmetry parameter $I=(N-Z)/A$.

Several investigations have been performed on the ground state properties
of (strongly) asymmetric nuclear matter, with large asymmetry parameter,
up to values as large as 0.8. In the extreme limit $I\to 1$ one obtains
neutron matter, which is obviously unbound as far as nuclear forces
are concerned, but deserves very interesting applications in the
description of neutron stars, which are bound by gravitational force,
but can be considered as a unique realization of this ideal system.

The equation of state of asymmetric nuclear matter contains a symmetry
term which depends upon the asymmetry parameter and the density of the
system; the equilibrium conditions are crucially determined by the
NN interaction: various effective forces can produce quite different
behaviours with $\rho$ in the potential energy contribution
to the symmetry energy.\cite{Ditoro1}

A study of non--equilibrium properties of asymmetric nuclear matter
has been carried out by Di Toro {\it et al.}\cite{Ditoro2} starting from
two Vlasov equations, for proton and neutron liquids, coupled through
the mean field. The unstable growth of density perturbations is
influenced by the initial asymmetry: in neutron rich nuclear matter
the formation of larger fragments, in which isospin
symmetry is restored, is favoured.

At low (subsaturation) densities the system develops collective
dynamical instabilities of isovector nature, the so--called 
{\it spinodal decomposition}: it corresponds to the growth of
density pertubations leading to a liquid--gas phase separation in
which symmetric nuclei coexist with a neutron gas. The instability
region (for example in the $(T,\rho)$ plane, turns out to be reduced
in strongly asymmetric systems.

This phenomenon can be tested in heavy ion reactions with radioactive
beams and could provide important information on large systems of
astrophysical interest.

\subsection{Neutron stars: structure and equation of state}

A large amount of information on neutron stars should be available in
the next few years from the new generation of X-- and $\gamma$--ray
satellites. In view of this, neutron stars are the subject of many
theoretical studies, aimed to predict their structure on the basis of
the properties of dense matter. The equation of state (EOS) of neutron
stars covers a wide density range, from $\sim 10$~g/cm$^3$ in the
surface to several times nuclear matter saturation density
($\rho\equiv\rho_0\sim
2.8\times 10^{14}$g/cm$^3$) in the center of the star.

The core (interior part) of a neutron star is believed to consist of
asymmetric nuclear matter with a consistent fraction of leptons. At
ultra--high densities, matter might undergo a transition giving rise
to other exotic hadronic components, like hyperons, a $K^-$ condensate
or a deconfined phase of quark matter. This occurrence, in turn, could
critically influence the evolution of neutron stars and could
eventually lead to the formation of black holes.\cite{Bomb97}

Baldo {\it et al.} have derived the static properties of non--rotating
neutron stars in a conventional framework, using the microscopic EOS
for asymmetric nuclear matter, derived from the Brueckner--Bethe--Goldstone
many--body theory.\cite{Baldo97} They use the Argonne $V_{14}$ and the
Paris two--body interactions, implemented by the Urbana model for the
three--body force. The latter are incorporated into the Brueckner scheme
by reducing them to affective two--body, density dependent forces. The
parameters are adjusted to reproduce the empirical nuclear matter
saturation point.

The calculated EOS allows to compute masses and radii as a
function of the central density $n_0$. Assuming that a neutron star
is a spherically symmetric distribution of mass in hydrostatic
equilibrium, and neglecting the effects of rotations and magnetic field,
the equilibrium configurations are obtained by solving the Tolman--
Oppenheimer--Volkoff (TOV) equations for the total pressure and the
enclosed mass:
\bea
\frac{dP(r)}{dr} &=& -\frac{Gm(r)\rho(r)}{r^2}
\frac{\displaystyle{\left(1+\frac{P(r)}{c^2\rho(r)}\right)
\left(1+\frac{4\pi r^3 P(r)}{c^2 m(r)}\right)}}
{\displaystyle{\left(1-\frac{2G m(r)}{rc^2}\right)}}
\label{dpress}\\
\frac{dm(r)}{dr} &=& 4\pi r^2 \rho(r),
\label{dm}
\eea
where $G$ is the gravitational constant. The maximum mass configuration
obtained in ref.\cite{Baldo97} range from 1.8 to 2.13 solar masses,
while the corresponding radii vary from 8 to 10.6~Km. These values are
consistent with the present observations.

A challenging suggestion is found in a work of Bombaci\cite{Bomb98},
concerning the semiempirical mass--radius relation extracted for the
X--ray burst source 4U 1820--30: the latter cannot be reproduced by
neutron star models based on conventional EOS. It is thus suggested
that the onset of a phase transition to a $K^-$ condensate could explain,
with a suitable choice of the parameters, the observed mass--radius
relation.
     
\subsection{Neutron stars: the effects of superfluidity}

As already mentioned in the above subsections,
it is believed that the inner crust of a neutron star, at a density of
about $3\times 10^{11}$~g/cm$^3$ (corresponding to about
$10^{-3}\rho_0$), is a superfluid and
inhomogeneous system, consisting of a lattice of nuclei
immersed in a sea of neutrons and an approximately uniform
sea of electrons. Such a configuration persists up to roughly
$\rho_0/2$. At low temperatures, this system is superfluid with a
positive Fermi energy $\epsilon_F$: the estimate of the pairing gap
in this inhomogeneous medium has been carried out by Barranco
{\it et al.}\cite{Barranco1}. They solve the Hartree--Fock--Bogoliubov
(HFB) equations in a Wigner--Seitz cell, with a Woods--Saxon
potential; the gap is calculated self--consistently without altering
the single particle levels. Use is made of the Argonne two--body
interaction. 

The HFB method is suitable to describe the halo--nuclei and/or
neutron--rich nuclei.  Surface collective modes (phonons) mediate
the induced interaction producing Cooper pairs and a bosonic
condensate.

In a second work\cite{Barranco2} the quantum
calculation is compared to the LDA. Here the two--body interaction
is assumed to be a Gogny force and is included in the HFB
equations via the pairing field. 

It is found that the LDA leads to a spatial variation of the
gap near the surface of a nucleus, which is stronger than the one
obtained in the HFB calculation. This is caused by the neglect
of the proximity effects and the delocalized character of the
single--particle wave functions close to the fermi energy.

From the energy of the system,
\beq
<E> =\sum_q n_q E_q,
\eeq
where $n_q=(1+e^{E_q/T})^{-1}$ is the occupation number (at the
temperature $T$) for the quasi--particle state $q$, one can obtain
the {\it specific heat} of the system:
\beq
C_v=\frac{1}{V}\frac{\partial <E>}{\partial T}\, .
\label{specH}
\eeq
This quantity turns out to be very sensitive to the pairing gap,
and hence to the presence of the inhomogeneity induced by the
presence of the nuclei in the crust. 
The overestimate of the gap, as obtained in LDA, leads to a
specific heat of the system which is too large at low temperatures,
as compared with the quantal result. Incidentally, a reliable
estimate of the pairing gap is also important for the determination
of the cooling time of the neutron star.

In concluding this argument, it is worth mentioning that pairing
correlations have been studied within a relativistic mean field
approach based on a field theory of nucleons coupled to  neutral
($\sigma$ and $\omega$) and charged ($\rho$) mesons. The HF and
pairing fields are calculated in a self--consistent way\cite{Matera}.
The energy gap is the result of strong cancellations between the
scalar and vector components of the pairing field. It is found
that the pair amplitude vanishes beyond a certain value of momentum
of the paired nucleons; the estimated gap is in agreement with
non--relativistic calculations of this quantity.

\subsection{Neutron stars: massive quark matter}

While in the previous subsection the attention was focussed on the
surface region of a neutron star, where the low density advocates
nucleonic degrees of freedom, its high central density, up to
$5\div 10~\rho_0$, seems to imply the existence of a core of massive
quark matter. The latter can be described in a variety of different
models, among which we shall consider here the so--called Color
Dielectric Model (CDM), which has been successfully employed in
connection with the single nucleon properties (structure functions,
e.m. form factors) as well as for the nucleon--nucleon interaction.

The CDM entails confinement of the quarks in ordinary hadronic
matter but also allows to describe, in mean field approximation,
a phase transition to deconfined quark matter\cite{Drago1, Drago2,
Drago3}. The model is defined by the following Lagrangian:
\bea
{\cal L} &&= i{\bar\psi}\gamma^{\mu}\partial_{\mu}\psi
+\sum_{f=u,d}\frac{g_f}{f_{\pi}\chi}{\bar\psi}_f\left(\sigma +
i\gamma_5\tau\cdot\pi\right)\psi_f 
\nonumber\\
&& + \frac{g_s}{\chi}{\bar\psi}_s\psi_s 
+\frac{1}{2}\left(\partial_{\mu}\chi\right)^2 
-\frac{1}{2}{\cal M}^2\chi^2
\label{CDMlag}\\
&& +\frac{1}{2}\left(\partial_{\mu}\sigma\right)^2 
+\frac{1}{2}\left(\partial_{\mu}\pi\right)^2 -U(\sigma,\pi)
\nonumber
\eea
where $\chi$ is the color dielectric field and $U(\sigma,\pi)$ the
 ``mexican hat'' potential of the chiral sigma--model. This Lagrangian
describes a system of interacting $u$, $d$ and $s$ quarks, pions, 
sigmas and a scalar--isoscalar chiral singlet field $\chi$. The latter
is related to the fluctuations of the gluon condensate around its 
vacuum expectation value. The coupling constants are given by 
$g_{u,d}=g(f_\pi\pm \xi_3)$ and $g_s=(2f_K-f_\pi)$, where $f_\pi=93$~MeV 
and $f_K=113$~MeV are the pion and kaon decay constants, respectively, 
while $\xi_3=f_{K^{\pm}}-f_{K^0}=-0.75$~MeV. Hence the model contains
only two free parameters, $g$ and ${\cal M}$, which are fixed to the
values $g=0.023$~GeV, ${\cal M}=1.7$~GeV. 
Confinement is obtained via the effective quark masses, which diverge
outside the nucleon. 

The CDM is employed to describe the deconfined quark matter phase in
the neutron star, while the relativistic field theoretical model of 
Walecka is used for the hadronic phase. Applying Gibbs criteria to this
composite system, Drago {\it et al.}\cite{Drago2} find that the pure 
hadronic phase ends at $0.11$~fm$^{-3}$ while the mixed (quark and hadronic)
phase extends up to $0.31$~fm$^{-3}$.

According to this calculation, neutron stars with masses in the range
$1.3 < M/M_{\odot} <1.54$ and radii of about 9~Km are found. A neutron star
with total mass of $1.4~M_{\odot}$ will consist of a crust  of pure 
hadronic matter, a $\sim 1$~Km thick region of mixed phase and a core of
pure quark matter.

In a recent work\cite{Drago3} the EOS of quark matter based on the CDM
has been discussed, taking into account the effects associated with
finite temperatures. Both the evolution of neutron stars and the onset
of supernovae explosion crucially depend upon the EOS of matter at very
high densities and temperatures. At finite $T$ the EOS considered here
shows a decrease of the pressure and of the adiabatic index, leading to
a deconfinement transition for densities slightly larger than the one
corresponding to nuclear matter saturation. The presence of a mixed phase
region softens the EOS and could lead to a direct supernova explosion.
At larger densities the EOS is stiff enough to support a neutron star
compatible with observations.

\section{Quark matter}

Quantum Chromodynamics (QCD), the non--abelian theory of coloured quarks
and gluons, is currently accepted as the theory of strong interactions,
and its predictions have been tested in a variety of 
elementary particle reactions at large momentum transfers. 
Its behaviour in the high energy 
(or short distance)  limit approaches the one of a non--interacting
free field theory ({\it asymptotic freedom}), while at low energies (or
large length scales) QCD is a non--perturbative field theory,
quarks and gluons being permanently confined inside hadrons. 

At very high nuclear densities and/or  temperatures it is believed that
hadronic matter undergoes a phase transition to a deconfined phase of
quarks and gluons, the so--called  Quark Gluon Plasma (QGP). In
this new phase hadrons dissolve, strong interactions become very weak 
and an ideal colour--conducting plasma of quarks and gluons is formed. 
In the QGP the long--range colour force is Debye--screened due to collective
effects and the quarks can only interact via a short--range effective
potential.

During the last decade the attention of nuclear and particle physics
has been attracted by the possibility of producing this new state of 
matter in the laboratory, by means of ultra--relativistic heavy 
ion collisions. A large number of experiments have been carried
out at the AGS (Brookhaven) and at the SPS (CERN). Both laboratories
have planned considerable ``upgrading''  of the existing facilities, 
the so--called RHIC (Relativistic Heavy Ion
Collider), which is going into operation in 1999 and the LHC (Large
Hadron Collider), whose program will be partly dedicated to high energy
particle physics and partly to heavy ion collisions.
The most recent measurements refer to S+U reactions at 200~GeV/A and
Pb+Pb at 158~GeV/A, reaching an (estimated) energy density of
$2\div 5$~GeV/fm$^3$. 
                                 
Many aspects of the transition from hadrons to deconfined quark matter,
among which the order of the phase transition and the nature of 
experimentally observable signatures, are still under debate.
Recently the NA50 experiment at CERN\cite{NA50} has shown an anomalous
suppression of the J/$\Psi$ survival probability measured
in Pb on Pb collisions\cite{Satz0}, which might be ascribed to the
formation of QGP at some stage of the collision. 
Other interesting signals could be revealed by the dilepton spectra,
which are sensitive to the masses of vector mesons: the latter
can be significantly altered by the presence of the QGP, an argument
which is still widely discussed.

\subsection{Quark deconfinement}

At high densities and temperatures, the
 interaction between quarks and gluons dresses their propagation, 
so that gluons develop an effective mass, which in
turn produces screening of the long--range colour--electric forces. 
For example, the free gluon propagator
$D_0(\om,k)\sim (\om^2-k^2)^{-1}$ is modified by summing an infinite 
chain of one--loop  insertions. One obtains\cite{Kapust2} the following 
longitudinal  gluon propagator 
\beq
D_L(\om,k)=\frac{1}{k^2\eps_L(\om,k)}
\label{gluonl}
\eeq
$D_L$ being a scalar function of the variables 
$\omega=k^0$ and $k=|\vec k|$ and $\eps_{L}$ the so--called 
colour dielectric function. From explicit calculations, one can
show that the static ($\om=0$) longitudinal colour fields are 
screened, being 
\beq
D_L(0,k)= \frac{1}{k^2\eps_L(0,k)}= \frac{1}{k^2+g^2 T^2}
\equiv \frac{1}{k^2+\lambda^{-2}_D}
\label{static}
\eeq
which defines the Debye length $\lambda_D=(gT)^{-1}$.

The above mentioned features of the gluon propagator entail several 
important consequences, among which we focus here on the fact that
the potential between two static colour charges (e.g. two heavy 
quarks) is {\it screened} in the quark--gluon plasma phase. Indeed the 
Fourier transform of $D_L(0,k)$, eq.(\ref{static}), yields the potential
\beq
V_{Q{\bar Q}}(r)\simeq \frac{1}{r}e^{-r/\lambda_D}
\label{Vscreen}
\eeq
with screening length $\lambda_D\simeq 0.4$~fm at $T=250$~MeV. 
This screening of the long range colour forces is believed to be 
 at the origin of quark deconfinement in the high temperature phase. 
It also leads to the disappearence of the bound states of a charmed 
quark pairs $(c{\bar c})$ in the QGP\cite{Matsui}. 

%%%%%%%%%%%%%%%%%%%%%%%%   Figure 10  %%%%%%%%%%%%%%%%%%%%%%%%%%

\begin{figure}
\centerline{
\epsfxsize=9cm
\epsfysize=9.5cm \epsfbox{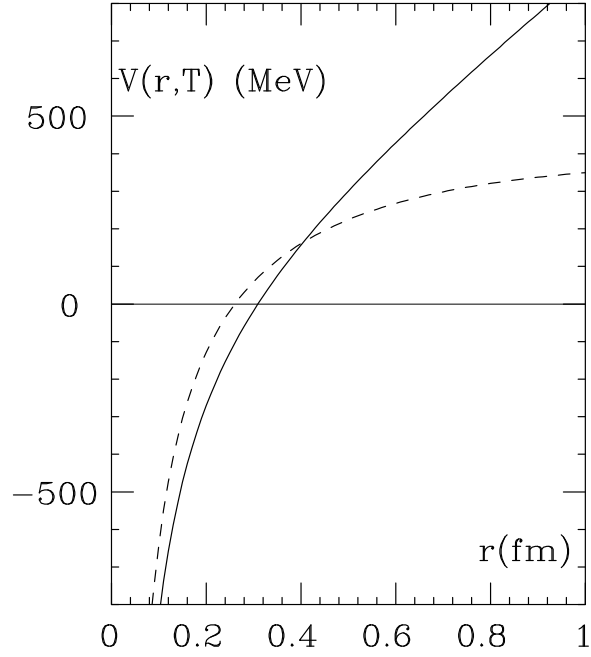}  
}
\caption[ ]{ The quark--quark potential, eq.(\ref{potsatz}) at $T=250^{\circ}$ 
(dashed line) and $T=0$ (continuous line).  }
\label{fig10}
\end{figure}

Fig.~10 illustrates the deconfinement effect of a thermodynamical enviroment 
of interacting light quarks and gluons on the interquark potential employed
in ref.\cite{KMS} to investigate the deconfinement conditions for heavy 
quarkonia.:
\beq
V(r,T)=\frac{\sigma}{\mu(T)}\left(1-e^{\mu(T)r}\right)
-\frac{\alpha}{r}e^{-\mu(T)r}
\label{potsatz}
\eeq
where $\mu(T)=1/\lambda_D(T)$. The potential (\ref{potsatz}) reduces
to the Cornell potential in the zero temperature limit (we remind that
$\lambda_D\to\infty$ for $T\to 0$); in addition to the one gluon exchange
contribution ($1/r$) this potential also contains a phenomenological
linear binding term, as suggested by lattice QCD calculations for the
interaction between static (heavy) colour charges. 

It would be important to find some ``QCD inspired'' model, suitable 
to describe both the confined (hadronic) and the deconfined (plasma) 
phases of a system of {\it many} quarks. This is obviously not trivial 
to achieve, due to the non--perturbative nature of the quark--quark 
interaction in the confined phase and also to the unavoidable 
approximations required to describe a many--body system, even when 
the interactions are known.

In a recent work Alberico {\it et al.}\cite{Piotr} have developed 
a three--dimensional model for quark matter with a {\it density 
dependent} quark--quark (confining) potential,  which allows 
to describe a sort of deconfinement transition as the system evolves 
from a low density assembly of bound structures to a high density 
free Fermi gas of quarks. Different confining potentials are considered,
some of which successfully utilized in hadron spectroscopy, like the
Cornell one. 

We find that a proper treatment of the many--body correlations 
induced by the medium is essential to disentangle the different nature 
of the two (hadronic and deconfined) phases of the system. For this 
purpose the ground state energy per particle and the pair 
correlation function are investigated. The latter  can be obtained 
from the expectation value of the two--body density operator:
\beq
g(r) =\frac{N(N-1)}{\rho^2}
<\Psi|\rho_2(|{\vec r}_1-{\vec r}_2|)|\Psi>, 
\label{defgr}
\eeq
where $|\Psi>$ is the exact (normalized) ground state of the system.
For a system of strongly correlated pairs $g(r)$ will show up a
well localized peak at small values of $r$, while the Fermi gas
pair correlation function has a fairly constant behaviour, with
the exception of small $r$ values, where the Pauli principle prevents
particles to be close to each other.

As an example, Fig.~11 shows the pair correlation function derived from
the Bethe--Goldstone wave functions of the screened Cornell potential:
\beq
V_{Cornell}(r,\rho) = \left(-\frac{a}{r} + br +K\right)e^{-c\rho r}
\label{Vcornel}
\eeq
where $a,b,c$ and $K$ are constants, while $\rho$ is the density of
the system.  The free Fermi gas two--body correlation function is 
recovered for $\rho\simeq 0.6$~fm$^{-3}$, in agreement with the value
of the ``transition density'' which can be obtained from the 
corresponding equation of state.

%%%%%%%%%%%%%%%%%%%   Figure 11   %%%%%%%%%%%%%%%%%%%%%%%%%%%%%

\begin{figure}
\centerline{
\epsfxsize=9.5cm
\epsfysize=9cm \epsfbox{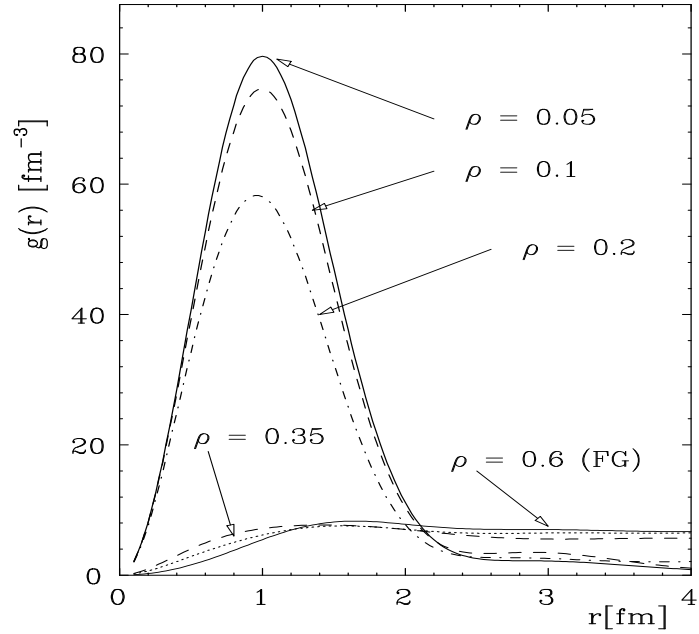}  
}
\caption[ ]{The pair correlation function $g(r)$ obtained with the
potential (\ref{Vcornel}) is displayed as a function of the
relative interquark distance at various densities: 
$\rho=0.05$~fm$^{-3}$ (thick solid line), $\rho=0.1$~fm$^{-3}$ (dashed 
line), $\rho=0.2$~fm$^{-3}$ (dot--dashed line),
$\rho=0.35$~fm$^{-3}$ (dashed line)  and $\rho=0.6$~fm$^{-3}$ 
(dotted line). For the last density value the Fermi gas correlation 
function is also shown (thin solid line).    }
\label{fig11}
\end{figure}

\subsection{Finite size effects in the QGP}

Before concluding this Section it is worth mentioning that one of the
major difficulties in the analysis of the relativistic heavy ion 
collisions stems from the numerous uncertainties in the thermostatistical
evolution of the system. Among the various steps, the hadronization 
process is crucial in determining the final states measured in the 
detectors. 

Brink {\it et al.}\cite{Brink} have studied the hadronization of a 
plasma, in order to establish whether  the final particles are 
evaporated from droplets of hadronic matter inside the plasma or
rather from the whole volume of the plasma itself. The two situations
can be distinguished on the basis of the entropy density of the system.

Let us assume that at equilibrium (close to the critical temperature 
$T\sim T_c$) the volume of the plasma ($V_{tot}$) is filled by $N$ droplets 
of QGP (each occupying a volume $V_{glob}$) and a gas of relativistic 
pions. For a fixed total volume and energy density, the calculated 
entropy density of the systems turns out to be larger when the plasma 
fills a big, unique glob rather than several globs of smaller radius. 
This is partially illustrated in Fig.~12. Finite size corrections are
taken into account in the density of levels which enters into the 
calculation of the entropy. Fluctuations with respect to the number of
globs are small, hence there is no tendency for a large glob to break
spontaneously into smaller ones.

%%%%%%%%%%%%%%%%%%%%   Figure 12    %%%%%%%%%%%%%%%%%%%%%%%%%%%%%%%%%%%%%%

\begin{figure}
\centerline{
\epsfxsize=11cm
\epsfysize=8cm \epsfbox{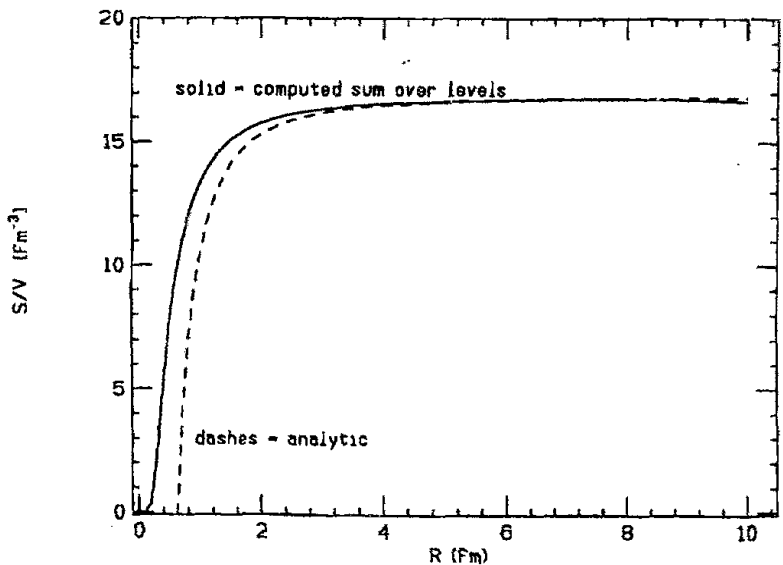}  
}
\caption[ ]{ Entropy density for a gas of quarks and gluons as a 
function of the glob radius $R$. The temperature is fixed to be $T=200$~MeV. }
\label{fig12}
\end{figure}

This argument concludes the tour which was foreseen in the introduction,
showing how the various many--body techniques can be applied to strongly
interacting systems, either of nucleons or of quarks, in an attempt
to provide a unified microscopic description of the intriguing, complex
structure of nuclear systems.

\section*{Acknowledgments}
I would like to thank Dr. A. de Pace and Prof. A. Molinari for
enlightening discussions  and to acknowledge their valuable help.

%%%%%%%%%%%%%%%%%%%%%%%%%%%%%%%%%%%%%%%%%%%%%%%%%%%%%%%%%%%%%

\end{document}